\documentclass[prl,aps,superscriptaddress,twocolumn]{revtex4-1}
\usepackage{amsfonts,amssymb,amscd,amsthm}
\usepackage{graphicx}
\usepackage{threeparttable}
\usepackage{multirow}
\usepackage{mathrsfs}
\usepackage[intlimits]{amsmath}
\usepackage{tabularx}
\usepackage{xcolor}
\usepackage[colorlinks, citecolor=red]{hyperref}
\usepackage{lineno}
\usepackage{etoolbox}
\usepackage{color,soul}
\usepackage{physics}
\usepackage{ulem}

\begin{document}

\title{Multiplexed double-transmon coupler scheme in scalable superconducting quantum processor}

\author{Tianqi Cai}
    \thanks{These authors contributed equally to this work.}
    \affiliation{Tencent Quantum Laboratory, Tencent, Shenzhen, Guangdong, China}
    
\author{Chitong Chen}
    \thanks{These authors contributed equally to this work.}
    \affiliation{T Lab, Shenzhen, Guangdong, China}

\author{Kunliang Bu} 
    \thanks{These authors contributed equally to this work.}
    \affiliation{Tencent Quantum Laboratory, Tencent, Shenzhen, Guangdong, China}

\author{Sainan Huai}
    \affiliation{Tencent Quantum Laboratory, Tencent, Shenzhen, Guangdong, China}

\author{Xiaopei Yang}
    \affiliation{Tencent Quantum Laboratory, Tencent, Shenzhen, Guangdong, China}

\author{Zhiwen Zong}
    \affiliation{Tencent Quantum Laboratory, Tencent, Shenzhen, Guangdong, China}
    
\author{Yuan Li}
    \affiliation{Tencent Quantum Laboratory, Tencent, Shenzhen, Guangdong, China}

\author{Zhenxing Zhang}
    \email{zzxht3@gmail.com}
    \affiliation{Tencent Quantum Laboratory, Tencent, Shenzhen, Guangdong, China}
 
\author{Yi-Cong Zheng}
    \email{yicongzheng@tencent.com}
    \affiliation{Tencent Quantum Laboratory, Tencent, Shenzhen, Guangdong, China}

\author{Shengyu Zhang}
    \email{shengyzhang@tencent.com}
    \affiliation{Tencent Quantum Laboratory, Tencent, Shenzhen, Guangdong, China}

\begin{abstract}

Precise control of superconducting qubits is essential for advancing both quantum simulation and quantum error correction. Recently, transmon qubit systems employing the single-transmon coupler (STC) scheme have demonstrated high-fidelity single- and two-qubit gate operations by dynamically tuning the effective coupling between qubits. However, the integration of STCs increases the number of control lines, thereby posing a significant bottleneck for chip routing and scalability. To address this challenge, we propose a robust control line multiplexing scheme based on a double-transmon coupler (DTC) architecture, which enables shared coupler control lines to substantially reduce wiring complexity. Moreover, we experimentally verify that this multiplexed configuration efficiently suppresses undesirable static $ZZ$ coupling while maintaining accurate control over two-qubit gate operations. We further demonstrate the feasibility of the architecture through two distinct gate implementations: a fast coupler $Z$-control-based CZ gate and a parametric iSWAP gate. To validate the practical applicability of this multiplexing approach in quantum circuits, we prepare Bell and three-qubit GHZ states using the proposed scheme with fidelity exceeding 99\% and 96\%, respectively. This multiplexed DTC architecture offers significant potential to minimize wiring overhead in two-dimensional qubit arrays, thereby greatly enhancing the scalability of superconducting quantum processors.
\end{abstract}

\maketitle

\section{Introduction}

Substantial progress has been made in large-scale quantum chip integration within the field of superconducting quantum computing in recent years~\cite{arute2019quantum, wu2021strong, google2023suppressing, cao2023generation, kim2023evidence, xu2023digital}. Systems based on transmon qubits~\cite{koch2007charge} have successfully integrated over 100 qubits, establishing a critical foundation for practical quantum computing applications~\cite{google2025quantum, gao2025establishing}. Despite these achievements, the development of scalable quantum processors remains hindered by two primary challenges. The first challenge stems from the geometric scaling of the control lines, where the number of control lines increases linearly with qubit counts. This scaling inevitably leads to an increase in noise channels, degradation of qubit coherence, and heightened demands on cryogenic cooling infrastructure~\cite{krinner2019engineering}. The second challenge concerns inter-qubit coupling crosstalk~\cite{valles2025optimizing}: residual transverse couplings induce frequency shifts in qubit transitions, while persistent $ZZ$ couplings cause undesired phase accumulation in the $\ket{11}$ computational basis states~\cite{barends2016digitized, kandala2019error, zhao2022quantum}.

Considerable efforts have been devoted to addressing these challenges. Approaches include frequency-division multiplexing of readout lines~\cite{jerger2012frequency, jeffrey2014fast, lecocq2021control}, multiplexed control lines shared among multiple qubits~\cite{asaad2016independent, manenti2021full, shi2023multiplexed, zhao2023baseband, zhao2024multiplexed, matsuda2025selective}, and the use of fixed-frequency qubits to eliminate $Z$ control lines for frequency tuning~\cite{kim2023evidence}. Additionally, the introduction of the single-transmon coupler (STC)~\cite{yan2018tunable} has enabled mitigation of inter-qubit residual coupling crosstalk through dynamic tuning of effective qubit-qubit coupling strengths~\cite{collodo2020implementation, xu2020high, foxen2020demonstrating, zhao2021suppression}. However, each STC requires an additional control line to adjust the coupler frequency. Although several alternative coupler designs have been proposed~\cite{mundada2019suppression, stehlik2021tunable, heunisch2023tunable, liang2023tunable, huber2024parametric}, most face similar challenges related to control line overhead. Notably, the recently proposed double-transmon coupler (DTC)~\cite{goto2022double, kubo2023fast, campbell2023modular, kubo2024high, li2024realization, li2025capacitively} has demonstrated superior control over effective coupling strengths and enabled high-fidelity controlled-Z (CZ) gate implementation in two-qubit systems~\cite{li2024realization}, although its scalability in multi-qubit configurations requires further validation.

In this work, we present a robust control-line multiplexed DTC scheme that achieves an optimal balance between control-line simplification and enhanced coupling precision. We systematically analyze and investigate the $ZZ$ coupling characteristics within this architecture, both theoretically and experimentally, confirming the feasibility of parallel single-qubit gate operations. The measured single-qubit gate fidelities obtained via simultaneous randomized benchmarking (RB)~\cite{magesan2012efficient, gambetta2012characterization} are consistent with those from isolated operations. Furthermore, we introduce a practical configuration supporting both single-qubit gate implementation and a scalable CZ gate operation. Through experimental realization of CZ gates characterized by RB, we quantify the impact of neighboring spectator qubits on gate fidelity, thereby demonstrating the robustness of these multiplexed gate operations. The preparation of Bell states and three-qubit Greenberger-Horne-Zeilinger (GHZ) states further validates the scalability of this approach. Finally, we propose a parametric gate implementation framework to demonstrate compatibility with two-dimensional quantum chip architectures. These results establish the proposed architecture as a promising candidate for scalable superconducting quantum processors.

%--------------------------      Figure 1      --------------------------%
\begin{figure}[!htb]
\includegraphics{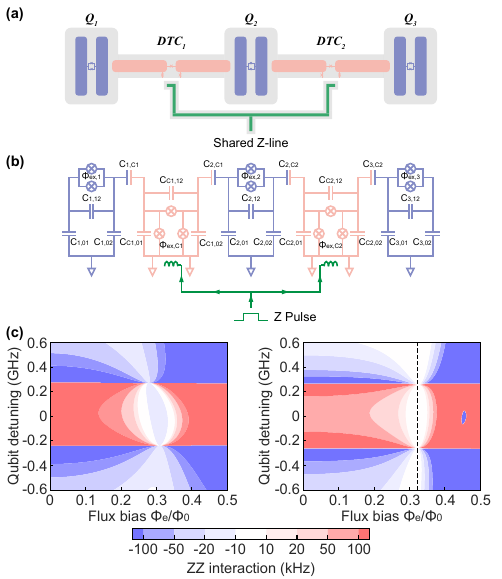}
\caption{\textbf{The multiplexed DTC scheme.} (a) Schematic diagram of a basic unit in the multiplexed DTC scheme, where three transmon qubits are coupled via two DTCs. The effective coupling strengths between qubits $Q_1$-$Q_2$ and $Q_2$-$Q_3$ are mediated by adjacent DTCs, which share a common $Z$ control line split into two branches on the chip. (b) Simplified circuit model corresponding to (a). Each DTC is represented as an integration of a fixed-frequency transmon qubit and a capacitively-shunted flux qubit (CSFQ)~\cite{li2024realization}. (c) Comparison of $ZZ$ coupling strengths in STC and DTC configurations. Based on our experimental circuit parameters, the STC exhibits significant variation in $ZZ$ coupling strength as a function of qubit detuning, whereas the DTC maintains a relatively stable $ZZ$ coupling. Moreover, the regime in which the $ZZ$ coupling remains below 10 kHz for the STC configuration is confined to a narrow crossing region, while the DTC sustains this low $ZZ$ coupling over a substantially broader range. This extended suppression window facilitates the practical implementation of control-line multiplexing within DTC architectures.}
\label{fig:Fig1}
\end{figure}
%--------------------------      Figure 1      --------------------------%
	
\section{Theoretical Model}

In this section, we present the theoretical model of our multiplexed DTC scheme. We begin by considering a multiplexed unit comprising three transmon qubits coupled via two DTCs, as illustrated in Fig.~\ref{fig:Fig1}(a). Each qubit is individually addressable through dedicated $XYZ$ control lines, while the two DTCs mediate the effective coupling strength between nearest-neighbor qubits via synchronized modulation supplied by a shared $Z$ control line. The simplified circuit schematic is depicted in Fig.~\ref{fig:Fig1}(b), where each DTC is represented as an integration of a fixed-frequency transmon qubit and a capacitively-shunted flux qubit (CSFQ)~\cite{li2024realization}. The system Hamiltonian describing a pair of neighboring qubits is given by
\begin{equation}\label{eq: H_DTC}
\begin{split}
H =& \sum_{i=1,2} \left(\omega_i \hat{a}_i^{\dag}\hat{a}_i + \frac{\eta_i}{2}\hat{a}_i^{\dag}\hat{a}_i^{\dag}\hat{a}_i\hat{a}_i\right)\\
& + \hat{H}_{p} + \hat{H}_{m}(\varphi_{ex})\\
& + g_{1p}(\hat{a}_1^{\dag}\hat{a}_p+\hat{a}_1\hat{a}_p^{\dag}) + g_{2p}(\hat{a}_2^{\dag}\hat{a}_p+\hat{a}_2\hat{a}_p^{\dag})\\
& + g_{1m}(\hat{a}_1^{\dag}\hat{a}_m+\hat{a}_1\hat{a}_m^{\dag}) - g_{2m}(\hat{a}_2^{\dag}\hat{a}_m+\hat{a}_2\hat{a}_m^{\dag}),\\
\end{split}
\end{equation}
where $\omega_i$ and $\eta_i$ are the frequency and anharmonicity of $Q_i$, $\hat{H}_p$ and $\hat{H}_m$ denote the internal Hamiltonians of the DTC $p$- and $m$-modes, $g_{ip}$ and $g_{im}$ indicate the coupling strength between $Q_i$ and the corresponding DTC modes, $\hat{a}^\dagger$ and $\hat{a}$ correspond to the creation and annihilation operators. In the dispersive regime $|\Delta_{ip(m)}| = |\omega_i - \omega_{p(m)}| \gg g_{ip(m)}$, the effective coupling of the system $g_\mathrm{eff}$ can be given as~\cite{li2024realization}
\begin{equation}\begin{aligned}
g_\mathrm{eff} = \frac{g_{1p}g_{2p}}{2}\left(\frac{1}{\Delta_{1p}}+\frac{1}{\Delta_{2p}}\right) - \frac{g_{1m}g_{2m}}{2}\left(\frac{1}{\Delta_{1m}}+\frac{1}{\Delta_{2m}}\right),
\end{aligned}\end{equation}
with $\Delta_{ip(m)} = \omega_i - \omega_{p(m)}$ denoting the detunings between each qubit and the DTC modes. For fixed qubit frequencies, the effective coupling strength $g_\mathrm{eff}$ can be modulated by adjusting the frequency of the $m$-mode $\omega_m$. In particular, when $\omega_m \approx \omega_p$, the $p$- and $m$-mode contributions can cancel, suppressing $g_\mathrm{eff}$ toward zero. In this regime, $g_\mathrm{eff}$ is relatively insensitive to modest qubit-frequency variations because $g_{ip} \approx g_{im}$, enabling real-time modulation of the effective coupling during circuit execution and thereby significantly reducing control overhead in scalable architectures.

%--------------------------      Figure 2      --------------------------%
\begin{figure*}[!htb]
\includegraphics{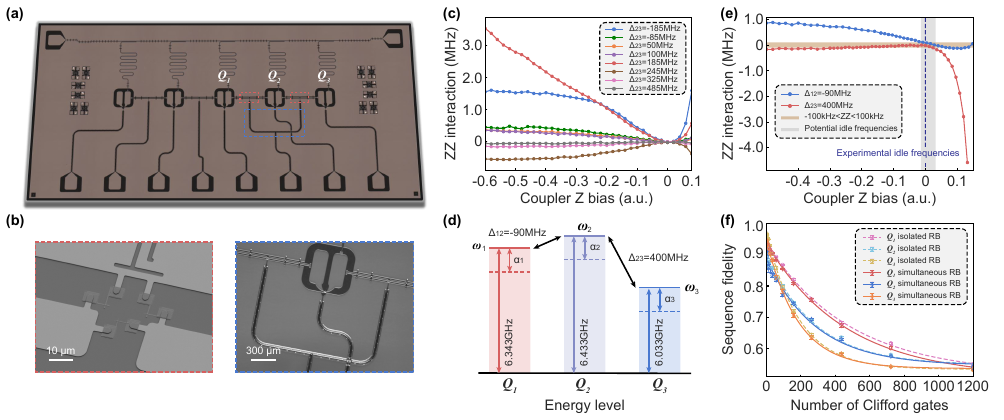}
\caption{\textbf{Chip configuration and $ZZ$ coupling.} (a) Optical micrograph of the superconducting quantum processor featuring five transmon qubits arranged in a one-dimensional chain architecture. The rightmost three qubits, $Q_i \ (i=1 \sim 3)$, are utilized to demonstrate the fundamental multiplexed unit, wherein the two DTCs coupling each qubit pair share a common $Z$ control line. (b) Enlarged views of the DTC and the multiplexed control lines. (c) Variation of $ZZ$ coupling as a function of qubit detuning and DTC $Z$ bias. Here, the frequency of $Q_2$ is held constant while the frequencies of $Q_3$ and the corresponding DTC are varied. The $ZZ$ suppression point shows negligible dependence on qubit detuning. (d) Idle frequency configuration of the three qubits $Q_i \ (i=1 \sim 3)$. The frequency detuning between $Q_1$-$Q_2$ is set within the straddling regime, whereas the detuning between $Q_2$-$Q_3$ lies outside this regime. (e) $ZZ$ coupling as a function of DTC $Z$ bias at the idle frequency configuration. The gray shaded region highlights the $Z$ bias range over which both qubit pairs maintain $ZZ$ coupling strengths below 100 kHz. (f) Comparison of gate fidelities obtained from simultaneous and isolated RB for qubits $Q_1$, $Q_2$ and $Q_3$. The consistent fidelities across simultaneous RB [$Q_1$: 99.91(1)\%; $Q_2$: 99.90(1)\%; $Q_3$: 99.86(1)\%] and isolated RB [$Q_1$: 99.92(2)\%; $Q_2$: 99.90(1)\%; $Q_3$: 99.87(1)\%] confirm effective suppression of $ZZ$ crosstalk within the multiplexed DTC scheme.}
\label{fig:Fig2}
\end{figure*}
%--------------------------      Figure 2      --------------------------%

We numerically characterize the static $ZZ$ interaction for both STC and DTC configurations; the results are summarized in Fig.~\ref{fig:Fig1}(c). Our analysis clearly indicates that the DTC architecture is markedly better suited to control‑line multiplexing, for two main reasons. First, in the STC configuration the $ZZ$ strength depends strongly on the detuning of neighboring qubits: when one qubit is held fixed and the other is tuned, residual $ZZ$ appears and produces frequency shifts that degrade both single-qubit and two-qubit gate fidelities. While active compensation—dynamically adjusting neighboring coupler control pulses—can keep $ZZ$ near zero~\cite{li2020tunable, gao2025establishing}, this requires applying flux pulses to all adjacent couplers whenever a single qubit is tuned, which incurs prohibitive control overhead given the large combinatorics of qubit‑frequency configurations during circuit execution. The commonly used workaround of increasing qubit detuning also has drawbacks: it forces qubits to traverse wider frequency ranges during non‑adiabatic two‑qubit gates, increasing exposure to spurious two‑level systems (TLSs)~\cite{klimov2018fluctuations, klimov2024Optimizing}, exacerbating pulse‑distortion errors~\cite{rol2020time,gao2025establishing, li2025high}, and ultimately reducing gate fidelity. Second, full suppression of $ZZ$ interaction in the STC design is typically achievable only within the straddling regime; outside that regime residual $ZZ$ persists despite increased detuning, leading to unwanted phase accumulation in the $\ket{11}$ state and further fidelity loss. By contrast, the DTC exhibits robust $ZZ$ suppression over a wide range of qubit frequencies: suppression points persist and a stable zero‑coupling bandwidth exists in which $ZZ$ remains below 10 KHz. These characteristics make the DTC architecture substantially more compatible with control‑line multiplexing, particularly for multi‑qubit systems that require simultaneous frequency tuning.

\section{Experimental Demonstration}

We implement the multiplexed DTC scheme on a superconducting quantum processor, comprising five frequency-tunable transmon qubits arranged in a one-dimensional chain. Each pair of neighboring qubits is coupled via a DTC, as illustrated in Fig.~\ref{fig:Fig2}(a). The three qubits on the right side, denoted $Q_i \ (i=1,2,3)$, serve to demonstrate the basic multiplexed unit, where the two DTCs responsible for modulation of qubit coupling share a common $Z$ control line, as shown in Fig.~\ref{fig:Fig2}(b). All control lines are encapsulated with fully-capped airbridges to ensure equipotential grounding and to minimize unwanted crosstalk~\cite{bu2025tantalum}. Each qubit is coupled to a common readout transmission line through individual readout resonators, enabling simultaneous single-shot measurements. Meanwhile, the readout frequencies are designed below the qubit frequencies to facilitate rapid qubit reset~\cite{reed2010fast}.

\subsection{Suppression of static $ZZ$ crosstalk}

We first verify that the DTC effectively suppresses $ZZ$ crosstalk between neighboring qubits with minimal sensitivity to their frequency detuning. To characterize the strength of the $ZZ$ interaction $\xi_{ZZ}$, we employ a Ramsey-type measurement protocol, which involves probing the frequency of the target qubit with the control qubit in either its ground or excited state. Taking qubit pair $Q_2-Q_3$ as an example, the frequency of $Q_2$ is kept constant, while the frequencies of $Q_3$ and the corresponding DTC are varied to observe the changes in $ZZ$ coupling, as shown in Fig.~\ref{fig:Fig2}(c). A key observation is that under near-zero $Z$ bias of the DTC, all curves converge closely at the same point, indicating that the position of the $ZZ$ suppression point remains essentially unchanged as the frequency difference between two qubits varies over a wide range. This behavior confirms that, unlike the STC, the intrinsic $ZZ$ suppression mechanism of the DTC renders the suppression point largely insensitive to the detuning of the qubit frequency\cite{goto2022double,campbell2023modular}.

We next demonstrate that two DTCs sharing a common $Z$ control line can simultaneously and effectively suppress $ZZ$ crosstalk between two qubit pairs. To achieve consistent $ZZ$ suppression points and enable high-fidelity two-qubit gate operations, precise optimization of DTC design parameters before device fabrication is critical. The key design specifications of the sample used in this study are detailed in the Supplementary Materials. To facilitate subsequent gate operations, the frequencies of the three qubits were carefully arranged in a $\Lambda$-type configuration. This frequency allocation mitigates leakage effects associated with multiplexed control lines, as illustrated in Fig.~\ref{fig:Fig2}(c). Specifically, the frequency detuning between $Q_1$ and $Q_2$ is $\Delta_{12}=-90$ MHz, while that between $Q_2$ and $Q_3$ is $\Delta_{23}=400$ MHz. In other words, one qubit pair’s frequency difference lies within the straddling regime, whereas the other lies outside it. Detailed device parameters are provided in the Supplementary Materials. In this frequency configuration, we performed a comprehensive characterization of the $ZZ$ interactions between $Q_1-Q_2$ and $Q_2-Q_3$, as shown in Fig.~\ref{fig:Fig2}(e). The results indicate that the $ZZ$ suppression points for both qubit pairs occur at approximately the same DTC frequency. However, due to fabrication tolerances and variations in the magnetic environment, slight deviations in the actual $ZZ$ suppression points are observed. These residual $ZZ$ interactions correspond to the remaining crosstalk within the multiplexed DTC scheme, as shown in the gray-shaded regions of Fig.~\ref{fig:Fig2}(e). Despite these deviations, the $ZZ$ values near the suppression points remain relatively low (<100 kHz). We anticipate that further improvements in cryogenic magnetic shielding and fabrication processes will help minimize these residual effects.

We utilize simultaneous RB to assess the effectiveness and robustness of $ZZ$ suppression within the multiplexed DTC scheme, as presented in Fig.~\ref{fig:Fig2}(f). Here, simultaneous RB is performed by executing RB sequences concurrently on $Q_1$, $Q_2$ and $Q_3$. The gate fidelities resulting from simultaneous RB [$Q_1$: 99.91(1)\%; $Q_2$: 99.90(1)\%; $Q_3$: 99.86(1)\%] closely match those obtained from isolated RB measurements [$Q_1$: 99.92(2)\%; $Q_2$: 99.90(1)\%; $Q_3$: 99.87(1)\%]. This close agreement demonstrates that with effective $ZZ$ suppression, adjacent qubits are well protected against crosstalk.

%--------------------------      Figure 3      --------------------------%
\begin{figure}[!htb]
\includegraphics{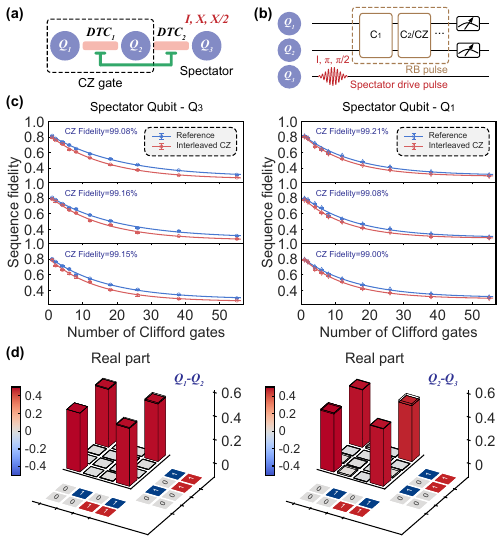}
\caption{\textbf{CZ gate and spectator qubit effects.} (a) Schematic diagram illustrating the investigation of the CZ gate and spectator qubit effects. After initializing all qubits in the $\ket{0}$ state, the spectator qubit is prepared in different states using the $I$, $X$ and $X/2$ operations to assess its impact on the error rate of the target CZ gate. (b) Pulse sequence for RB experiments. Following the application of drive pulses to the spectator qubit, the RB protocol is executed on the target qubit pair, which is subsequently measured. (c) CZ gate fidelities extracted from RB experiments. The left panel shows results with $Q_3$ as the spectator qubit, while the right panel corresponds to $Q_1$ as the spectator qubit. The top, middle, and bottom rows represent the outcomes when applying the $I$, $X$ and $X/2$ gates to the spectator qubit, respectively. The extracted CZ gate fidelities are 99.08\%, 99.16\%, and 99.15\% (left panel), and 99.21\%, 99.08\%, and 99.00\% (right panel). (d) QST density matrices for Bell states prepared on qubit pairs $Q_1$-$Q_2$ and $Q_2$-$Q_3$, achieving fidelities of 99.34\% and 99.57\%, respectively.}
\label{fig:Fig3}
\end{figure}
%--------------------------      Figure 3      --------------------------%

\subsection{Implementation of two-qubit CZ gate}

We next validate the multiplexed DTC scheme by implementing a two-qubit controlled-Z (CZ) gate. The DTC utilizes its $p$-mode and $m$-mode couplings with adjacent qubits to establish an effective two-qubit interaction . During the CZ gate operation, carefully engineered flux pulses with slowly changing waveforms are applied to each qubit to bring them into resonance under the condition $\omega_{11}=\omega_{20}$. Concurrently, flux pulses are applied to the two DTCs based on the shared $Z$ line, and following a controlled evolution period, a $\pi$ phase is accumulated, thereby realizing a standard diabatic CZ gate~\cite{strauch2003quantum, barends2019diabatic, foxen2020demonstrating}. Importantly, the flux pulses applied to the DTCs require precise optimization to achieve strong $XY$ coupling for the target qubit pair while simultaneously suppressing $ZZ$ coupling for the other pair. This balance is attainable through meticulous chip parameter design. In practice, the DTC-mediated $ZZ$ coupling exhibits a relatively broad low-coupling regime (as illustrated in Fig.~\ref{fig:Fig1}(c)) and can be made to feature several engineered “off” (near‑zero) points. By leveraging these characteristics and applying appropriate design strategies, CZ gate operations can be performed with negligible interference from $ZZ$ coupling with neighboring spectator qubits.

Experimentally, we implemented CZ gates on the qubit pairs $Q_1-Q_2$ and $Q_2-Q_3$, and evaluated the impact of $ZZ$ coupling from the corresponding spectator qubit, as illustrated in Fig.~\ref{fig:Fig3}(a). Taking the CZ gate between $Q_1-Q_2$ as an example, $Q_3$ was designated as the spectator qubit. We prepared $Q_3$ sequentially in the states $\ket{0}$, $\ket{1}$ and $\frac{1}{\sqrt{2}}(\ket{0}+\ket{1})$, while performing RB sequences on $Q_1-Q_2$, as shown in Fig.~\ref{fig:Fig3}(b). By measuring the CZ gate fidelity, we quantified the $ZZ$ crosstalk between $Q_2-Q_3$ and assessed the effectiveness of its suppression. The characterization results, presented in Fig.~\ref{fig:Fig3}(c), indicate that when $Q_3$ functions as the spectator qubit, the RB fidelities of the CZ gates applied to $Q_1-Q_2$ are 99.08\%, 99.16\%, and 99.15\% following the application of the identity ($I$), $X$, and $X/2$ gates to $Q_3$, respectively, with a maximum observed variation of only 0.08\%. Similarly, when $Q_1$ serves as the spectator qubit, the corresponding CZ gate fidelities on $Q_2-Q_3$ are 99.21\%, 99.08\%, and 99.00\%, exhibiting a maximum variation of 0.22\%. These findings confirm that the residual $ZZ$ coupling between neighboring qubits is effectively suppressed by the multiplexed DTC scheme, even during qubit frequency excursions, without requiring additional flux pulses on the couplers. This capability substantially reduces the influence of spectator qubits on adjacent CZ gate operations, highlighting the promising scalability of this architecture.

To further verify the entangling capability of the system, Bell states were generated on the qubit pairs $Q_1-Q_2$ and $Q_2-Q_3$ using the CZ gates. Quantum state tomography (QST)~\cite{neeley2010generation} was conducted to characterize and accurately quantify the state fidelities. The results, shown in Fig.~\ref{fig:Fig3}(d), demonstrate that the Bell states prepared on $Q_1-Q_2$ and $Q_2-Q_3$ achieved fidelities of 99.34\% and 99.57\%, respectively.

%--------------------------      Figure 4      --------------------------%
\begin{figure}[!htb]
\includegraphics{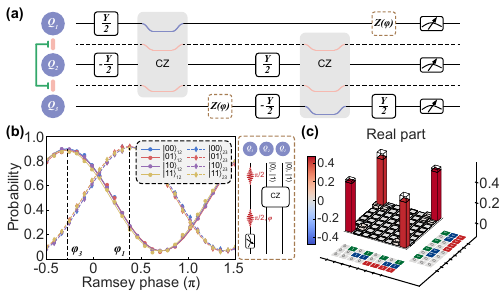}
\caption{\textbf{Generation of three-qubit GHZ state.} (a) Pulse sequence employed for the generation of three-qubit GHZ state. Here, $Z(\phi)$ denotes the virtual $Z$ gate used for phase compensation. (b) Phase calibration protocol for measuring single-qubit $Z$ phase shifts. The right panel illustrates the schematic of the Ramsey-type experiment designed to quantify the phase accumulated in the spectator qubit during CZ gate operations applied to the target qubit pair. The left panel presents the corresponding experimental data along with fitted curves, from which the phase offset is extracted. (c) QST density matrix of the generated three-qubit GHZ state, demonstrating a measured fidelity of 96.0\%.}
\label{fig:Fig4}
\end{figure}
%--------------------------      Figure 4      --------------------------%

\section{Discussions}

The preceding theoretical analysis and experimental results demonstrate that, even with a shared coupler $Z$ line, there exist sufficient degrees of freedom to maintain precise control over the system. In the following, we explore the applications of the multiplexed DTC scheme from two perspectives: first, the implementation of quantum circuits, and second, the investigation of scalable gate schemes and architectures.

\subsection{Generation of three-qubit GHZ state}

We begin by examining the execution of quantum circuits within the shared coupler $Z$ line configuration. We select a three-qubit GHZ state as the experimental target. The GHZ state not only generates maximal entanglement but also involves multi-qubit gate operations, making it an ideal testbed for comparing quantum circuits implemented under shared versus conventional non-shared frameworks. The experimental pulse sequence is depicted in Fig.~\ref{fig:Fig4}(a). In practice, because the $Z$ lines of the two DTCs are shared, performing a CZ gate on the qubit pair $Q_1-Q_2$ ($Q_2-Q_3$) simultaneously induces $Z$ modulation on the corresponding DTC for $Q_2-Q_3$ ($Q_1-Q_2$), resulting in a shift of the DTC frequency. Due to the relatively strong direct coupling between the qubits and the DTC, this frequency shift causes a slight detuning of qubit frequencies. Consequently, when executing the CZ gate on $Q_1-Q_2$ ($Q_2-Q_3$), the frequency of $Q_3$ ($Q_1$) experiences a minor shift, which accumulates an additional single-qubit $Z$ phase on $Q_3$ ($Q_1$). This effect requires careful experimental characterization and compensation. Importantly, this phenomenon highlights a key distinction between shared and non-shared $Z$ line configurations. In the latter, assuming negligible classical $Z$ crosstalk, two-qubit operations on any qubit pair do not induce phase accumulation of other qubits.

To calibrate these single-qubit $Z$ phase shifts, we perform Ramsey-type experiments~\cite{sung2021realization} on $Q_3$ ($Q_1$) after preparing the qubit pair $Q_1-Q_2$ ($Q_2-Q_3$) into states $|00\rangle$, $|01\rangle$, $|10\rangle$, $|11\rangle$, respectively. The corresponding measurement results are shown in Fig.~\ref{fig:Fig4}(b). By fitting the data, we extract the relevant phases and apply virtual $Z$ gates to compensate for the shifts~\cite{mckay2017efficient}. It is observed that the four curves of phases corresponding to different initial states for $Q_1-Q_2$ ($Q_2-Q_3$) are nearly overlapping, indicating that spectator qubit is well isolated from target qubits. Finally, QST procedure is employed to measure the fidelity of the final circuit state, yielding a three-qubit GHZ state fidelity of 96.0\%, as shown in Fig.~\ref{fig:Fig4}(c). These results further confirm the feasibility and scalability of the multiplexed DTC scheme.

\subsection{Scalable parametric gate schemes}

We proceed to investigate scalable gate schemes that accommodate an increased number of DTCs sharing a common $Z$ line. The diabatic CZ gate, as previously described, can be effectively implemented when two DTCs share a $Z$ line. However, as more DTCs share one control line, implementing fast $Z$-control-based CZ gates via couplers becomes increasingly challenging due to limited selectivity, which complicates the identification of precise operating points. To address this challenge, we propose an alternative scalable parametric gate scheme that, in principle, supports a further increase in the number of DTCs sharing a $Z$ line.

\begin{figure}[tb]
\includegraphics{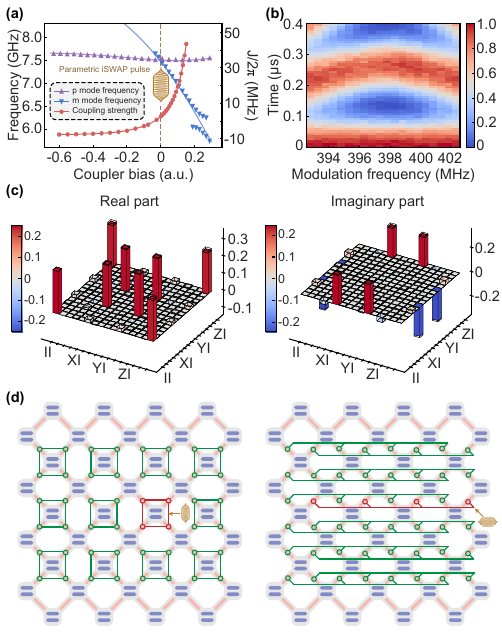}
\caption{\textbf{Realization of the parametric iSWAP gate.} (a) Conceptual schematic of the parametric iSWAP gate implementation using the multiplexed DTC scheme. The effective coupling strength $g_\mathrm{eff}$ is modulated to oscillate through the application of a parametric flux pulse on the DTCs. (b) Characterization of the effective coupling strength via chevron patterns. The qubits are initially prepared in the $\ket{10}$ (or $\ket{01}$) state, followed by fine-tuning of the qubit frequencies to achieve resonance. (c) Experimentally measured process matrix $\chi_\mathrm{exp}$ obtained from QPT of the parametric iSWAP gate, demonstrating a gate fidelity of 96.7\%. The left panel displays the real components, while the right panel shows the imaginary components. Solid black outlines correspond to the ideal gate. (d) Two potential approaches for implementing control line multiplexing within a two-dimensional square-lattice qubit architecture. The left panel depicts a central qubit whose four adjacent couplers are controlled via a single shared line, whereas the right panel illustrates multiplexing of control lines along each row.}
\label{fig:Fig5}
\end{figure}

Parametric gates utilize control waveforms at specific frequencies applied to either qubits or couplers, selectively driving the desired two-qubit transitions through frequency matching, thereby providing intrinsic selectivity~\cite{didier2018analytical, caldwell2018parametrically, li2022realization}. Here, we employ DTCs with shared $Z$ lines to realize parametric iSWAP gates of $Q_2-Q_3$ as a representative example, demonstrating the feasibility of this approach. As illustrated in Fig.~\ref{fig:Fig5}(a), the qubits are biased at the idle points shown in Fig.~\ref{fig:Fig2}(d), while the DTCs are positioned near their “off” points. By applying modulation at the energy difference between the $\ket{01}$ and $\ket{10}$ states of the two-qubit system $Q_2-Q_3$ on the DTCs, the corresponding iSWAP gate is activated, as depicted in the chevron patterns of Fig.~\ref{fig:Fig5}(b). At this operating point, the effective coupling $g_\mathrm{eff}$ between the qubits exhibits periodic oscillations near the zero-coupling point, described as
\begin{equation}\begin{aligned}
g_\mathrm{eff}(t) =& \frac{g_{1p}g_{2p}}{2}\left(\frac{1}{\Delta_{1p}}+\frac{1}{\Delta_{2p}}\right)\\
-& \frac{g_{1m}g_{2m}}{2}\left(\frac{1}{\Delta_{1m}(t)}+\frac{1}{\Delta_{2m}(t)}\right)\\
\approx& A_d\sin\left(\omega_d t+\phi_0\right),
\end{aligned}\end{equation}
where $A_d$ denotes the amplitude of $g_{\text{eff}}(t)$, $\omega_d$ represents the oscillation frequency of $g_{\text{eff}}(t)$, and $\phi_0$ indicates the phase. When $\omega_d = \Delta_{23}$, the parametric iSWAP gate can then be implemented. Quantum process tomography (QPT) was performed to characterize the gate, yielding the process matrix shown in Fig.~\ref{fig:Fig5}(c). The raw data indicate a process fidelity of 96.7\%.

%--------------------------      Figure 5      --------------------------%
%--------------------------      Figure 5      --------------------------%

Finally, by leveraging the intrinsic selectivity of parametric gates, it becomes possible to share coupler $Z$ control lines among a larger number of DTCs. We observe that the multiplexed DTC scheme is naturally compatible with the two-dimensional square-lattice qubit architecture. Figure~\ref{fig:Fig5}(d) illustrates two potential approaches for implementing such line sharing. The first approach, shown in the left panel of Fig.~\ref{fig:Fig5}(d), involves a central qubit whose four adjacent couplers are controlled via a single shared line. This configuration is particularly well suited for the surface code~\cite{fowler2012surface}. The second approach, depicted in the right panel of Fig.~\ref{fig:Fig5}(d), utilizes alternating control line multiplexing along each row, potentially enabling enhanced multiplexing efficiency.

In summary, we propose a novel multiplexed DTC scheme that concurrently mitigates routing congestion in control lines and suppresses undesirable $ZZ$ interactions among multiple qubit pairs. Through Randomized Benchmarking, we show that the fidelity of parallel-executed single-qubit gates remains comparable to that of individually executed ones. We further introduce an implementation strategy for non-adiabatic two-qubit CZ gates. The fidelity of the CZ gate demonstrates resilience to the quantum states of spectator qubits, underscoring the robustness of the proposed scheme. The high-fidelity preparation of Bell states and three-qubit GHZ states confirms the feasibility of the scheme for quantum circuit execution. Finally, we highlight that the proposed multiplexed DTC scheme, when integrated with parametric gates, is compatible with two-dimensional square-lattice qubit architectures. This compatibility facilitates practical implementation of surface codes and other quantum error correction protocols, positioning the scheme as a promising candidate for scalable multiplexed architectures in large-scale quantum computing platforms.

\section{Data Availability}
The data that support the findings of this study are available from the corresponding authors Z.X.Z., Y.C.Z. and S.Y.Z. upon request.

\section{Code Availability}
The code that support the simulations of this study are available from the corresponding authors Z.X.Z., Y.C.Z. and S.Y.Z. upon request.

\section{Acknowledgements:} 
We thank Fuming Liu, Guanglei Xi, Qiaonian Yu and Hualiang Zhang for supporting room-temperature electronics. 

\section{Author Contributions:}
Z.X.Z., Y.C.Z. conceived the experiment. Z.X.Z., X.P.Y. S.N.H and T.Q.C designed the device. K.L.B. fabricated the devices with assistance of Y.L., T.Q.C. and S.N.H.. Z.X.Z., T.Q.C. established the measurement setup and performed experimental measurements with assistance of Z.W.Z. and S.N.H.. C.T.C. performed the theoretical analyses and numerical simulations with assistance of Z.X.Z. and T.Q.C.. T.Q.C., C.T.C., Z.X.Z. wrote the manuscript with feedback from all authors. Y.C.Z., S.Y.Z. supervised the project. All authors contributed to the discussion of the results and development of the manuscript.

\section{Competing interests}
The authors declare no competing interests.

%\bibliographystyle{unsrt}
%\bibliography{Citation}

\begin{thebibliography}{10}

\bibitem{arute2019quantum}
Frank Arute, Kunal Arya, Ryan Babbush, Dave Bacon, Joseph~C Bardin, Rami
  Barends, Rupak Biswas, Sergio Boixo, Fernando~GSL Brandao, David~A Buell,
  et~al.
\newblock Quantum supremacy using a programmable superconducting processor.
\newblock {\em Nature}, 574(7779):505--510, 2019.

\bibitem{wu2021strong}
Yulin Wu, Wan-Su Bao, Sirui Cao, Fusheng Chen, Ming-Cheng Chen, Xiawei Chen,
  Tung-Hsun Chung, Hui Deng, Yajie Du, Daojin Fan, et~al.
\newblock Strong quantum computational advantage using a superconducting
  quantum processor.
\newblock {\em Physical review letters}, 127(18):180501, 2021.

\bibitem{google2023suppressing}
Google~Quantum AI.
\newblock Suppressing quantum errors by scaling a surface code logical qubit.
\newblock {\em Nature}, 614(7949):676--681, 2023.

\bibitem{cao2023generation}
Sirui Cao, Bujiao Wu, Fusheng Chen, Ming Gong, Yulin Wu, Yangsen Ye, Chen Zha,
  Haoran Qian, Chong Ying, Shaojun Guo, et~al.
\newblock Generation of genuine entanglement up to 51 superconducting qubits.
\newblock {\em Nature}, 619(7971):738--742, 2023.

\bibitem{kim2023evidence}
Youngseok Kim, Andrew Eddins, Sajant Anand, Ken~Xuan Wei, Ewout Van Den~Berg,
  Sami Rosenblatt, Hasan Nayfeh, Yantao Wu, Michael Zaletel, Kristan Temme,
  et~al.
\newblock Evidence for the utility of quantum computing before fault tolerance.
\newblock {\em Nature}, 618(7965):500--505, 2023.

\bibitem{xu2023digital}
Shibo Xu, Zheng-Zhi Sun, Ke~Wang, Liang Xiang, Zehang Bao, Zitian Zhu, Fanhao
  Shen, Zixuan Song, Pengfei Zhang, Wenhui Ren, et~al.
\newblock Digital simulation of projective non-abelian anyons with 68
  superconducting qubits.
\newblock {\em Chinese Physics Letters}, 40(6):060301, 2023.

\bibitem{koch2007charge}
Jens Koch, Terri~M Yu, Jay Gambetta, Andrew~A Houck, David~I Schuster, Johannes
  Majer, Alexandre Blais, Michel~H Devoret, Steven~M Girvin, and Robert~J
  Schoelkopf.
\newblock Charge-insensitive qubit design derived from the cooper pair box.
\newblock {\em Physical Review A—Atomic, Molecular, and Optical Physics},
  76(4):042319, 2007.

\bibitem{google2025quantum}
Google~Quantum AI.
\newblock Quantum error correction below the surface code threshold.
\newblock {\em Nature}, 638(8052):920--926, 2025.

\bibitem{gao2025establishing}
Dongxin Gao, Daojin Fan, Chen Zha, Jiahao Bei, Guoqing Cai, Jianbin Cai, Sirui
  Cao, Fusheng Chen, Jiang Chen, Kefu Chen, et~al.
\newblock Establishing a new benchmark in quantum computational advantage with
  105-qubit zuchongzhi 3.0 processor.
\newblock {\em Physical Review Letters}, 134(9):090601, 2025.

\bibitem{krinner2019engineering}
Sebastian Krinner, Simon Storz, Philipp Kurpiers, Paul Magnard, Johannes
  Heinsoo, Raphael Keller, Janis Luetolf, Christopher Eichler, and Andreas
  Wallraff.
\newblock Engineering cryogenic setups for 100-qubit scale superconducting
  circuit systems.
\newblock {\em EPJ Quantum Technology}, 6(1):2, 2019.

\bibitem{valles2025optimizing}
S~Vall{\'e}s-Sanclemente, THF Vroomans, TR~van Abswoude, F~Brulleman,
  T~Stavenga, SLM van~der Meer, Y~Xin, A~Lawrence, V~Singh, MA~Rol, et~al.
\newblock Optimizing the frequency positioning of tunable couplers in a circuit
  qed processor to mitigate spectator effects on quantum operations.
\newblock {\em arXiv preprint arXiv:2503.13225}, 2025.

\bibitem{barends2016digitized}
Rami Barends, Alireza Shabani, Lucas Lamata, Julian Kelly, Antonio Mezzacapo,
  U~Las Heras, Ryan Babbush, Austin~G Fowler, Brooks Campbell, Yu~Chen, et~al.
\newblock Digitized adiabatic quantum computing with a superconducting circuit.
\newblock {\em Nature}, 534(7606):222--226, 2016.

\bibitem{kandala2019error}
Abhinav Kandala, Kristan Temme, Antonio~D C{\'o}rcoles, Antonio Mezzacapo,
  Jerry~M Chow, and Jay~M Gambetta.
\newblock Error mitigation extends the computational reach of a noisy quantum
  processor.
\newblock {\em Nature}, 567(7749):491--495, 2019.

\bibitem{zhao2022quantum}
Peng Zhao, Kehuan Linghu, Zhiyuan Li, Peng Xu, Ruixia Wang, Guangming Xue,
  Yirong Jin, and Haifeng Yu.
\newblock Quantum crosstalk analysis for simultaneous gate operations on
  superconducting qubits.
\newblock {\em PRX quantum}, 3(2):020301, 2022.

\bibitem{jerger2012frequency}
Markus Jerger, Stefano Poletto, Pascal Macha, Uwe H{\"u}bner, Evgeni
  Il’ichev, and Alexey~V Ustinov.
\newblock Frequency division multiplexing readout and simultaneous manipulation
  of an array of flux qubits.
\newblock {\em Applied Physics Letters}, 101(4), 2012.

\bibitem{jeffrey2014fast}
Evan Jeffrey, Daniel Sank, JY~Mutus, TC~White, J~Kelly, R~Barends, Y~Chen,
  Z~Chen, B~Chiaro, A~Dunsworth, et~al.
\newblock Fast accurate state measurement with superconducting qubits.
\newblock {\em Physical review letters}, 112(19):190504, 2014.

\bibitem{lecocq2021control}
Florent Lecocq, Franklyn Quinlan, Katarina Cicak, Jose Aumentado, SA~Diddams,
  and JD~Teufel.
\newblock Control and readout of a superconducting qubit using a photonic link.
\newblock {\em Nature}, 591(7851):575--579, 2021.

\bibitem{asaad2016independent}
Serwan Asaad, Christian Dickel, Nathan~K Langford, Stefano Poletto, Alessandro
  Bruno, Michiel~Adriaan Rol, Duije Deurloo, and Leonardo DiCarlo.
\newblock Independent, extensible control of same-frequency superconducting
  qubits by selective broadcasting.
\newblock {\em npj Quantum Information}, 2(1):1--7, 2016.

\bibitem{manenti2021full}
Riccardo Manenti, Eyob~A Sete, Angela~Q Chen, Shobhan Kulshreshtha, Jen-Hao
  Yeh, Feyza Oruc, Andrew Bestwick, Mark Field, Keith Jackson, and Stefano
  Poletto.
\newblock Full control of superconducting qubits with combined on-chip
  microwave and flux lines.
\newblock {\em Applied Physics Letters}, 119(14), 2021.

\bibitem{shi2023multiplexed}
Pan Shi, Jiahao Yuan, Fei Yan, and Haifeng Yu.
\newblock Multiplexed control scheme for scalable quantum information
  processing with superconducting qubits.
\newblock {\em arXiv preprint arXiv:2312.06911}, 2023.

\bibitem{zhao2023baseband}
Peng Zhao, Ruixia Wang, Meng-Jun Hu, Teng Ma, Peng Xu, Yirong Jin, and Haifeng
  Yu.
\newblock Baseband control of superconducting qubits with shared microwave
  drives.
\newblock {\em Physical Review Applied}, 19(5):054050, 2023.

\bibitem{zhao2024multiplexed}
Peng Zhao.
\newblock A multiplexed control architecture for superconducting qubits with
  row-column addressing.
\newblock {\em arXiv preprint arXiv:2403.03717}, 2024.

\bibitem{matsuda2025selective}
R~Matsuda, R~Ohira, T~Sumida, H~Shiomi, A~Machino, S~Morisaka, K~Koike,
  T~Miyoshi, Y~Kurimoto, Y~Sugita, et~al.
\newblock Selective excitation of superconducting qubits with a shared control
  line through pulse shaping.
\newblock {\em arXiv preprint arXiv:2501.10710}, 2025.

\bibitem{yan2018tunable}
Fei Yan, Philip Krantz, Youngkyu Sung, Morten Kjaergaard, Daniel~L Campbell,
  Terry~P Orlando, Simon Gustavsson, and William~D Oliver.
\newblock Tunable coupling scheme for implementing high-fidelity two-qubit
  gates.
\newblock {\em Physical Review Applied}, 10(5):054062, 2018.

\bibitem{collodo2020implementation}
Michele~C Collodo, Johannes Herrmann, Nathan Lacroix, Christian~Kraglund
  Andersen, Ants Remm, Stefania Lazar, Jean-Claude Besse, Theo Walter, Andreas
  Wallraff, and Christopher Eichler.
\newblock Implementation of conditional phase gates based on tunable zz
  interactions.
\newblock {\em Physical review letters}, 125(24):240502, 2020.

\bibitem{xu2020high}
Yuan Xu, Ji~Chu, Jiahao Yuan, Jiawei Qiu, Yuxuan Zhou, Libo Zhang, Xinsheng
  Tan, Yang Yu, Song Liu, Jian Li, et~al.
\newblock High-fidelity, high-scalability two-qubit gate scheme for
  superconducting qubits.
\newblock {\em Physical review letters}, 125(24):240503, 2020.

\bibitem{foxen2020demonstrating}
Brooks Foxen, Charles Neill, Andrew Dunsworth, Pedram Roushan, Ben Chiaro,
  Anthony Megrant, Julian Kelly, Zijun Chen, Kevin Satzinger, Rami Barends,
  et~al.
\newblock Demonstrating a continuous set of two-qubit gates for near-term
  quantum algorithms.
\newblock {\em Physical Review Letters}, 125(12):120504, 2020.

\bibitem{zhao2021suppression}
Peng Zhao, Dong Lan, Peng Xu, Guangming Xue, Mace Blank, Xinsheng Tan, Haifeng
  Yu, and Yang Yu.
\newblock Suppression of static zz interaction in an all-transmon quantum
  processor.
\newblock {\em Physical Review Applied}, 16(2):024037, 2021.

\bibitem{mundada2019suppression}
Pranav Mundada, Gengyan Zhang, Thomas Hazard, and Andrew Houck.
\newblock Suppression of qubit crosstalk in a tunable coupling superconducting
  circuit.
\newblock {\em Physical Review Applied}, 12(5):054023, 2019.

\bibitem{stehlik2021tunable}
J~Stehlik, DM~Zajac, DL~Underwood, T~Phung, J~Blair, S~Carnevale, D~Klaus,
  GA~Keefe, A~Carniol, Muir Kumph, et~al.
\newblock Tunable coupling architecture for fixed-frequency transmon
  superconducting qubits.
\newblock {\em Physical review letters}, 127(8):080505, 2021.

\bibitem{heunisch2023tunable}
Lukas Heunisch, Christopher Eichler, and Michael~J Hartmann.
\newblock Tunable coupler to fully decouple and maximally localize
  superconducting qubits.
\newblock {\em Physical Review Applied}, 20(6):064037, 2023.

\bibitem{liang2023tunable}
Gui-Han Liang, Xiao-Hui Song, Cheng-Lin Deng, Xu-Yang Gu, Yu~Yan, Zheng-Yang
  Mei, Si-Lu Zhao, Yi-Zhou Bu, Yong-Xi Xiao, Yi-Han Yu, et~al.
\newblock Tunable-coupling architectures with capacitively connecting pads for
  large-scale superconducting multiqubit processors.
\newblock {\em Physical Review Applied}, 20(4):044028, 2023.

\bibitem{huber2024parametric}
GBP Huber, FA~Roy, L~Koch, I~Tsitsilin, J~Schirk, NJ~Glaser, N~Bruckmoser,
  C~Schweizer, J~Romeiro, G~Krylov, et~al.
\newblock Parametric multi-element coupling architecture for coherent and
  dissipative control of superconducting qubits.
\newblock {\em arXiv preprint arXiv:2403.02203}, 2024.

\bibitem{goto2022double}
Hayato Goto.
\newblock Double-transmon coupler: Fast two-qubit gate with no residual
  coupling for highly detuned superconducting qubits.
\newblock {\em Physical review applied}, 18(3):034038, 2022.

\bibitem{kubo2023fast}
Kentaro Kubo and Hayato Goto.
\newblock Fast parametric two-qubit gate for highly detuned fixed-frequency
  superconducting qubits using a double-transmon coupler.
\newblock {\em Applied Physics Letters}, 122(6), 2023.

\bibitem{campbell2023modular}
Daniel~L Campbell, Archana Kamal, Leonardo Ranzani, Michael Senatore, and
  Matthew~D LaHaye.
\newblock Modular tunable coupler for superconducting circuits.
\newblock {\em Physical Review Applied}, 19(6):064043, 2023.

\bibitem{kubo2024high}
Kentaro Kubo, Yinghao Ho, and Hayato Goto.
\newblock High-performance multiqubit system with double-transmon couplers:
  Toward scalable superconducting quantum computers.
\newblock {\em Physical Review Applied}, 22(2):024057, 2024.

\bibitem{li2024realization}
Rui Li, Kentaro Kubo, Yinghao Ho, Zhiguang Yan, Yasunobu Nakamura, and Hayato
  Goto.
\newblock Realization of high-fidelity cz gate based on a double-transmon
  coupler.
\newblock {\em Physical Review X}, 14(4):041050, 2024.

\bibitem{li2025capacitively}
Rui Li, Kentaro Kubo, Yinghao Ho, Zhiguang Yan, Shinichi Inoue, Yasunobu
  Nakamura, and Hayato Goto.
\newblock Capacitively shunted double-transmon coupler realizing bias-free
  idling and high-fidelity cz gate.
\newblock {\em arXiv preprint arXiv:2503.03053}, 2025.

\bibitem{magesan2012efficient}
Easwar Magesan, Jay~M Gambetta, Blake~R Johnson, Colm~A Ryan, Jerry~M Chow,
  Seth~T Merkel, Marcus~P Da~Silva, George~A Keefe, Mary~B Rothwell, Thomas~A
  Ohki, et~al.
\newblock Efficient measurement of quantum gate error by interleaved randomized
  benchmarking.
\newblock {\em Physical review letters}, 109(8):080505, 2012.

\bibitem{gambetta2012characterization}
Jay~M Gambetta, Antonio~D C{\'o}rcoles, Seth~T Merkel, Blake~R Johnson, John~A
  Smolin, Jerry~M Chow, Colm~A Ryan, Chad Rigetti, Stefano Poletto, Thomas~A
  Ohki, et~al.
\newblock Characterization of addressability by simultaneous randomized
  benchmarking.
\newblock {\em Physical review letters}, 109(24):240504, 2012.

\bibitem{li2020tunable}
X~Li, T~Cai, H~Yan, Z~Wang, X~Pan, Y~Ma, W~Cai, J~Han, Z~Hua, X~Han, et~al.
\newblock Tunable coupler for realizing a controlled-phase gate with
  dynamically decoupled regime in a superconducting circuit.
\newblock {\em Physical Review Applied}, 14(2):024070, 2020.

\bibitem{klimov2018fluctuations}
Paul~V Klimov, Julian Kelly, Zijun Chen, Matthew Neeley, Anthony Megrant, Brian
  Burkett, Rami Barends, Kunal Arya, Ben Chiaro, Yu~Chen, et~al.
\newblock Fluctuations of energy-relaxation times in superconducting qubits.
\newblock {\em Physical review letters}, 121(9):090502, 2018.

\bibitem{klimov2024Optimizing}
Paul~V. Klimov, Andreas Bengtsson, Chris Quintana, Alexandre Bourassa, Sabrina
  Hong, Andrew Dunsworth, Kevin~J. Satzinger, William~P. Livingston, Volodymyr
  Sivak, and Murphy~Yuezhen Niu.
\newblock Optimizing quantum gates towards the scale of logical qubits.
\newblock {\em Nature Communications}, 15(1), 2024.

\bibitem{rol2020time}
Michiel~A Rol, Livio Ciorciaro, Filip~K Malinowski, Brian~M Tarasinski,
  Ramiro~E Sagastizabal, Cornelis~Christiaan Bultink, Yves Salathe, Niels
  Haandb{\ae}k, Jan Sedivy, and Leonardo DiCarlo.
\newblock Time-domain characterization and correction of on-chip distortion of
  control pulses in a quantum processor.
\newblock {\em Applied Physics Letters}, 116(5), 2020.

\bibitem{li2025high}
Tian-Ming Li, Jia-Chi Zhang, Bing-Jie Chen, Kaixuan Huang, Hao-Tian Liu,
  Yong-Xi Xiao, Cheng-Lin Deng, Gui-Han Liang, Chi-Tong Chen, Yu~Liu, et~al.
\newblock High-precision pulse calibration of tunable couplers for
  high-fidelity two-qubit gates in superconducting quantum processors.
\newblock {\em Physical Review Applied}, 23(2):024059, 2025.

\bibitem{bu2025tantalum}
Kunliang Bu, Sainan Huai, Zhenxing Zhang, Dengfeng Li, Yuan Li, Jingjing Hu,
  Xiaopei Yang, Maochun Dai, Tianqi Cai, Yi-Cong Zheng, et~al.
\newblock Tantalum airbridges for scalable superconducting quantum processors.
\newblock {\em npj Quantum Information}, 11(1):17, 2025.

\bibitem{reed2010fast}
Matthew~D Reed, Blake~R Johnson, Andrew~A Houck, Leonardo DiCarlo, Jerry~M
  Chow, David~I Schuster, Luigi Frunzio, and Robert~J Schoelkopf.
\newblock Fast reset and suppressing spontaneous emission of a superconducting
  qubit.
\newblock {\em Applied Physics Letters}, 96(20), 2010.

\bibitem{strauch2003quantum}
Frederick~W Strauch, Philip~R Johnson, Alex~J Dragt, CJ~Lobb, JR~Anderson, and
  FC~Wellstood.
\newblock Quantum logic gates for coupled superconducting phase qubits.
\newblock {\em Physical review letters}, 91(16):167005, 2003.

\bibitem{barends2019diabatic}
Rami Barends, CM~Quintana, AG~Petukhov, Yu~Chen, Dvir Kafri, Kostyantyn
  Kechedzhi, Roberto Collins, Ofer Naaman, Sergio Boixo, F~Arute, et~al.
\newblock Diabatic gates for frequency-tunable superconducting qubits.
\newblock {\em Physical review letters}, 123(21):210501, 2019.

\bibitem{neeley2010generation}
Matthew Neeley, Radoslaw~C Bialczak, M~Lenander, Erik Lucero, Matteo
  Mariantoni, AD~O’connell, D~Sank, H~Wang, M~Weides, J~Wenner, et~al.
\newblock Generation of three-qubit entangled states using superconducting
  phase qubits.
\newblock {\em Nature}, 467(7315):570--573, 2010.

\bibitem{sung2021realization}
Youngkyu Sung, Leon Ding, Jochen Braum{\"u}ller, Antti Veps{\"a}l{\"a}inen,
  Bharath Kannan, Morten Kjaergaard, Ami Greene, Gabriel~O Samach, Chris
  McNally, David Kim, et~al.
\newblock Realization of high-fidelity cz and zz-free iswap gates with a
  tunable coupler.
\newblock {\em Physical Review X}, 11(2):021058, 2021.

\bibitem{mckay2017efficient}
David~C McKay, Christopher~J Wood, Sarah Sheldon, Jerry~M Chow, and Jay~M
  Gambetta.
\newblock Efficient z gates for quantum computing.
\newblock {\em Physical Review A}, 96(2):022330, 2017.

\bibitem{didier2018analytical}
Nicolas Didier, Eyob~A Sete, Marcus~P da~Silva, and Chad Rigetti.
\newblock Analytical modeling of parametrically modulated transmon qubits.
\newblock {\em Physical Review A}, 97(2):022330, 2018.

\bibitem{caldwell2018parametrically}
SA~Caldwell, N~Didier, CA~Ryan, EA~Sete, A~Hudson, P~Karalekas, R~Manenti,
  MP~da~Silva, R~Sinclair, E~Acala, et~al.
\newblock Parametrically activated entangling gates using transmon qubits.
\newblock {\em Physical Review Applied}, 10(3):034050, 2018.

\bibitem{li2022realization}
Shaowei Li, Daojin Fan, Ming Gong, Yangsen Ye, Xiawei Chen, Yulin Wu, Huijie
  Guan, Hui Deng, Hao Rong, He-Liang Huang, et~al.
\newblock Realization of fast all-microwave controlled-z gates with a tunable
  coupler.
\newblock {\em Chinese Physics Letters}, 39(3):030302, 2022.

\bibitem{fowler2012surface}
Austin~G Fowler, Matteo Mariantoni, John~M Martinis, and Andrew~N Cleland.
\newblock Surface codes: Towards practical large-scale quantum computation.
\newblock {\em Physical Review A—Atomic, Molecular, and Optical Physics},
  86(3):032324, 2012.

\end{thebibliography}

\clearpage

\end{document}

\title{Supplementary Materials: Multiplexed double-transmon coupler scheme in scalable superconducting quantum processor}

\author{Tianqi Cai}
    \thanks{These authors contributed equally to this work.}
    \affiliation{Tencent Quantum Laboratory, Tencent, Shenzhen, Guangdong, China}
    
\author{Chitong Chen}
    \thanks{These authors contributed equally to this work.}
    \affiliation{T Lab, Shenzhen, Guangdong, China}

\author{Kunliang Bu} 
    \thanks{These authors contributed equally to this work.}
    \affiliation{Tencent Quantum Laboratory, Tencent, Shenzhen, Guangdong, China}

\author{Sainan Huai}
    \affiliation{Tencent Quantum Laboratory, Tencent, Shenzhen, Guangdong, China}

\author{Xiaopei Yang}
    \affiliation{Tencent Quantum Laboratory, Tencent, Shenzhen, Guangdong, China}

\author{Zhiwen Zong}
    \affiliation{Tencent Quantum Laboratory, Tencent, Shenzhen, Guangdong, China}
    
\author{Yuan Li}
    \affiliation{Tencent Quantum Laboratory, Tencent, Shenzhen, Guangdong, China}

\author{Zhenxing Zhang}
    \email{zzxht3@gmail.com}
    \affiliation{Tencent Quantum Laboratory, Tencent, Shenzhen, Guangdong, China}
 
\author{Yi-Cong Zheng}
    \email{yicongzheng@tencent.com}
    \affiliation{Tencent Quantum Laboratory, Tencent, Shenzhen, Guangdong, China}

\author{Shengyu Zhang}
    \email{shengyzhang@tencent.com}
    \affiliation{Tencent Quantum Laboratory, Tencent, Shenzhen, Guangdong, China}

\maketitle

\tableofcontents

\newpage

\section{Theory}

\subsection{Theoretical model}

%--------------------------      Figure S1      --------------------------
\begin{figure*}[htb]
\includegraphics{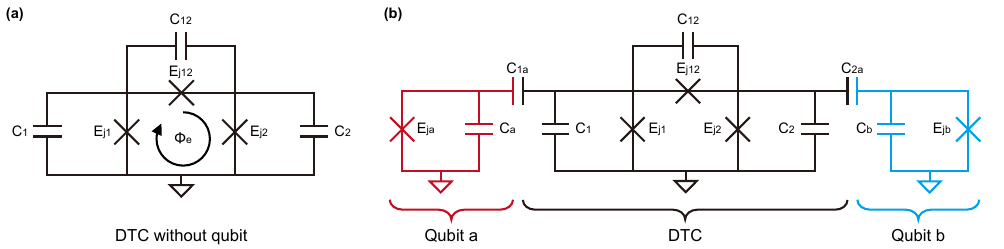}
\caption{\textbf{Theoretical model.} The circuit model for DTC (a) without data qubit, and (b) with data qubit.}
\label{fig:FigS1}
\end{figure*}
%--------------------------      Figure S1      --------------------------

In this section, we briefly describe the theory of double-transmon coupler (DTC)~\cite{li2024realization}. We first derive the toy model of DTC without the data qubits. The circuit of DTC is shown in Fig.~\ref{fig:FigS1}(a). 
The Lagrangian of DTC can be written as
\begin{equation}
\begin{aligned}
\mathcal{L} &= \frac{1}{2}\left(\frac{\Phi_0}{2\pi}\right)^2\left[C_1\dot{\phi}^2_1+C_2\dot{\phi}^2_2+C_{12}(\dot{\phi}_2-\dot{\phi}_1)^2\right]\\
&\quad+E_{j1}\cos(\phi_1)+E_{j2}\cos(\phi_2)\\
&\quad+E_{j12}\cos(\phi_2-\phi_1+\phi_e),
\end{aligned}
\end{equation}
where $\Phi_0 = \frac{h}{2e}$ is the flux quantum, $\phi_i$ ($i=1, \, 2$) is the node flux operator and $\phi_e$ is the external flux penetrating through the DTC. Assuming the symmetric design, $C_1=C_2=C_c$, $E_{j1}=E_{j2}=E_j$, and donating $E_{j12}=\alpha E_j$ with $\alpha < 1$, the Lagrangian becomes
\begin{equation}
\begin{aligned}
\mathcal{L} &= \left(\frac{\Phi_0}{2\pi}\right)^2\left[C_c\dot{\phi}^2_p+(C_c+2C_{12})\dot{\phi}_m^2\right]\\
&+2E_{j}\cos\phi_p\cos\phi_m+\alpha E_j\cos(\phi_m-\phi_e),
\end{aligned}
\end{equation}
with $\phi_{p,m}\equiv(\phi_2 \pm \phi_1)/2$. The potential energy is then given as $V=-2E_j\cos\phi_p\cos\phi_m-\alpha E_j\cos(2\phi_m+\phi_e)$, where $\alpha$ is about 0.27 in our design. Therefore, it can be further simplified as~\cite{li2024realization}
\begin{equation}
\begin{aligned}
V&=-2E_j\cos\phi_p -2E_j\cos\phi_m\\
&\quad-\alpha E_j(2\phi_m+\phi_e) -\frac{E_j}{2}\phi_p^2\phi_m^2,
\end{aligned}
\end{equation}
and here we omit the constant value without loss of generality. Using the Cooper-pair number operators $n_{p,m}=-i\partial/\partial\phi_{p,m}$, the Hamiltonian is expressed as 
\begin{equation}
\label{eq:dtc_ham}
\begin{aligned}
\mathcal{H} &= 4 E_{cp} n_p^2 + 4 E_{cm} n_m^2 - 2E_j\cos\phi_p\\
&\quad-2E_j\cos\phi_m - \alpha E_j\cos(2\phi_m+\phi_e) - \frac{E_j}{2}\phi_p^2\phi_m^2,
\end{aligned}
\end{equation}
where $E_{cp} = e^2/(4C_c)$, $E_{cm} = e^2/(4C_c+8C_{12})$. From Eq.~(\ref{eq:dtc_ham}), each DTC can be modeled as a composite element comprising a fixed‑frequency, transmon‑like $p$‑mode and a flux‑tunable CSFQ‑like $m$‑mode~\cite{yan2016flux}, with a direct inter‑mode coupling term that hybridizes their dynamics. 

Next we consider a circuit consisting of two transmon qubits coupled via a DTC, as illustrated in Fig.~\ref{fig:FigS1}(b). For clarity, we analyze the grounded qubit configuration here; the derivation for the floating qubit case discussed in the main text is similar. Under the symmetric assumptions $C_a=C_b=C_q$, $C_{1a}=C_{2b}=C_g$, the Lagrangian of the system can be written as
\begin{equation}
\begin{aligned}
\mathcal{L} &= K - V,\\
K &=\left(\frac{\Phi_0}{2\pi}\right)^2\bigg[\frac{C_a}{2}\dot{\phi}_a^2+\frac{C_b}{2}\dot{\phi}_b^2 + C_c\dot{\phi}_p^2+(C_c+2C_{12})\dot{\phi}_m^2\\
&\quad+\frac{C_{1a}}{2}(\dot{\phi}_a-\dot{\phi}_p+\dot{\phi}_m)^2+\frac{C_{2b}}{2}(\dot{\phi}_b-\dot{\phi}_p-\dot{\phi}_m)^2\bigg]\\
&=\left(\frac{\Phi_0}{2\pi}\right)^2\bigg[\frac{C_q}{2}\dot{\phi}_a^2+\frac{C_q}{2}\dot{\phi}_b^2 + C_c\dot{\phi}_p^2+(C_c+2C_{12})\dot{\phi}_m^2\\
&\quad+\frac{C_{g}}{2}(\dot{\phi}_a-\dot{\phi}_p+\dot{\phi}_m)^2+\frac{C_{g}}{2}(\dot{\phi}_b-\dot{\phi}_p-\dot{\phi}_m)^2\bigg],\\
V &=-E_{ja}\cos\phi_a - E_{jb}\cos\phi_b - 2E_j\cos\phi_p\cos\phi_m\\
&\quad-\alpha E_j\cos(2\phi_m+\phi_e)\\
&\approx-E_{ja}\cos\phi_a - E_{jb}\cos\phi_b - 2E_j\cos\phi_p-2E_j\cos\phi_m\\
&\quad-\alpha E_j\cos(2\phi_m+\phi_e)-\frac{E_j}{2}\phi_p^2\phi_m^2,
\end{aligned}
\end{equation}
where $K$ is the kinetic energy and $V$ is the potential energy. The kinetic energy can be rewritten in matrix form $K=\frac{1}{2}\bm{\dot{\Phi}}^T \bm{C}\bm{\dot{\Phi}}$, where $\dot{\bm{\Phi}}=\frac{\Phi_0}{2\pi}[\dot{\phi}_a, \dot{\phi}_b, \dot{\phi}_p, \dot{\phi}_m]^T$, and $\bm{C}$ can be represented as
\begin{equation}
\bm{C} = 
\begin{bmatrix}
 C_g+C_q & 0 & -C_g & C_g \\
 0 & C_g+C_q & -C_g & -C_g \\
 -C_g & -C_g & 2 C_c+2 C_g & 0 \\
 C_g & -C_g & 0 & 4 C_{12}+2 C_g+2 C_g \\
\end{bmatrix}
\end{equation}
The canonical conjugate variables are then defined as $\bm{Q}=\frac{\partial\mathcal{L}}{\partial \dot{\bm{\Phi}}} = \bm{C}\dot{\bm{\Phi}}$, and the Hamiltonian can then be expressed as
\begin{equation}
\label{eq:sys_ham_mat}
\begin{aligned}
\mathcal{H} &= \bm{Q}^T(\bm{C}^{-1})^{T}\bm{Q} -\mathcal{L} = 
\bm{Q}^T(\bm{C}^{-1})\bm{Q} -\mathcal{L}\\
&=2e^2 \bm{n}^T (\bm{C}^{-1}) \bm{n} + V,
\end{aligned}
\end{equation}
where $\bm{n} = \frac{1}{2e}\bm{Q}$. The inversion of the capacitance matrix $\bm{C}^{-1}$ can be calculated, and in the first order of the coupling capacitance $C_g$, we obtain the inversion matrix
\begin{widetext}
\begin{equation}
\bm{C}^{-1} \approx
\begin{bmatrix}
\frac{1}{C_q+C_g} & 0 & \frac{C_g}{2(C_g+C_q)(C_g+C_c)} & -\frac{C_g}{(C_q+C_g)(4C_{12}+2C_c+2C_g)}\\
0 & \frac{1}{C_q+C_g} & \frac{C_g}{2(C_g+Cq)(C_g+C_c)} & \frac{C_g}{(C_q+C_g)(4C_{12}+2C_c+2C_g)}\\
\frac{C_g}{2(C_g+C_q)(C_g+C_c)} & \frac{C_g}{2(C_g+C_q)(C_g+C_c)} & \frac{1}{2C_c+2C_g} & 0\\
-\frac{C_g}{(C_q+C_g)(4C_{12}+2C_c+2C_g)} & \frac{C_g}{(C_q+C_g)(4C_{12}+2C_c+2C_g))} & 0 & \frac{1}{2C_c+4C_{12} + 2 C_g}
\end{bmatrix}
\end{equation}
\end{widetext}
Expanding Eq.~(\ref{eq:sys_ham_mat}), the Hamiltonian can be written as
\begin{widetext}
\begin{equation}
\label{eq:sys_ham_full}
\begin{aligned}
\mathcal{H} &= \frac{2e^2}{C_q+C_g}n_a^2 -E_{ja}\cos\phi_a
 +\frac{2e^2}{C_q+C_g} n_b^2 -E_{jb}\cos\phi_b\\
&\quad + \frac{e^2}{C_c+C_g} n_p^2 -2E_j\cos\phi_p -\frac{E_j}{2}\phi_p^2\phi_m^2\\
&\quad \frac{e^2}{C_c+2C_{12}+C_g} n_m^2 -2E_j\cos\phi_m-\alpha E_j\cos(2\phi_m+\phi_e)\\
&\quad +\frac{2e^2C_g}{(C_c+C_g)(C_q+C_g)}(n_a+n_b)n_p + \frac{2e^2C_g}{(C_q+C_g)(C_c+2C_{12}+C_g)} (n_b-n_a)n_m. \\
\end{aligned}
\end{equation}    
\end{widetext}
The coupling term between qubit a and the $m$‑mode in Eq.~(\ref{eq:sys_ham_full}) gives rise to an effective negative interaction, which is crucial for suppressing or canceling the coupling between qubit a and qubit b in the DTC system.

Finally, the coupling strength $g_{jp}$, $g_{jm}$ ($j=a, \, b$) between qubits and the DTC can be derived as
\begin{equation}
\label{eq:mode_coupling}
\begin{aligned}
g_{jp} &= \sqrt{\omega_{j}\omega_p}\frac{C_g}{\sqrt{2(C_c+C_g)(C_q+C_g)}}\\
g_{jm} &=\sqrt{\omega_{j}\omega_m}\frac{C_g}{\sqrt{2(C_c+C_g+2C_{12})(C_q+C_g)}}. \\
\end{aligned}
\end{equation}
Following the Schrieffer-Wolff transformation (SWT)~\cite{yan2018tunable}, we obtain the effective coupling strength between qubits:
\begin{equation}
\label{eq:effective_coupling}
g_\mathrm{eff} = \frac{g_{ap}g_{bp}}{2}\left(\frac{1}{\Delta_{ap}}+\frac{1}{\Delta_{bp}}\right) - \frac{g_{am}g_{bm}}{2}\left(\frac{1}{\Delta_{am}}+\frac{1}{\Delta_{bm}}\right),
\end{equation}
where $\Delta_{jp} = \omega_j - \omega_p$, $\Delta_{jm} = \omega_j - \omega_m$ ($j=a, \, b$). Since $C_{12} \ll C_c$ in our design, the coupling strength between qubit and $p$-mode is approximately equal to that of the $m$-mode, i.e., $g_{am}\approx g_{ap}$, $g_{bm}\approx g_{bp}$. As a result, the contributions from the two modes interfere destructively, and the effective interqubit coupling $g_\mathrm{eff}$ vanishes when the $p$‑ and $m$‑mode frequencies are matched ($\omega_p\approx\omega_m$).

\subsection{Design parameters}
\begin{table}[htb]
\caption{Design parameters for our quantum processor. All capacitive symbols are defined in Fig.~1(b) of the main text with $j=1, \, 2, \, c$. $E_{jq}$ represents the Josephson energy for the qubit; $E_{j1}$, $E_{j2}$ represent the Josephson energies for the DTC defined in Fig.~\ref{fig:FigS1}, and $\alpha=\frac{E_{j12}}{E_{j}}$ with $E_{j1}=E_{j2}=E_{j}$.}
\begin{tabular*}{\linewidth}{@{\extracolsep{\fill}}cccc}
\hline
\hline
 &{Qubit} &{DTC} &{Qubit-DTC coupling} \\
\hline
$C_{j, 01}$ (fF) &{$82.9$} &{$84.6$} &{$\,$} \\
$C_{j, 02}$ (fF) &{$78.6$} &{$84.6$} &{$\,$} \\
$C_{j, 12}$ (fF) &{$32.9$} &{$4.0$} &{$\,$} \\
$C_{1, c}$ (fF) &{$\,$} &{$\,$} &{$11.3$} \\
$C_{2, c}$ (fF) &{$\,$} &{$\,$} &{$11.3$} \\
$E_{jq}/h$ (GHz) &{$23.1$} &{$\,$} &{$\,$} \\
$E_{j1}, E_{j2}/h$ (GHz) &{$\,$} &{$36.0$} &{$\,$} \\
$\alpha$ &{$\,$} &{$0.27$} &{$\,$}\\
\hline
\end{tabular*}
\label{table:DeviceParameters}
\end{table}

\section{Device and Fabrication}

Our quantum processor consists of five transmon qubits interconnected via four tunable DTCs. In experiment, we focus on demonstrating the system using three qubits \blue{\sout{on}} $Q_1$, $Q_2$, and $Q_3$ as illustrated in Fig.~2(a) of the main text. The computational qubits are arranged in a linear chain configuration, with each qubit equipped with independent $XYZ$ control lines. The $XY$ and $Z$ control signals are combined through a cryogenic diplexer to optimize signal routing~\cite{manenti2021full}. Each DTC incorporates a dual-junction SQUID design, allowing frequency tuning via an externally applied magnetic flux. Among the four DTCs, two are controlled by dedicated $Z$ lines, while the remaining two (coupling $Q_1-Q_2$ and $Q_2-Q_3$) share common control lines. Furthermore, each qubit is coupled to an individual readout resonator, enabling dispersive readout of the qubit states. The DTCs do not possess dedicated resonator and their characterization is performed indirectly through measurements of the coupled qubits.

The device was fabricated on a 430 $\mu$m-thick sapphire substrate coated with a 200 nm-thick $\alpha$-tantalum film. The base circuit pattern was defined using photolithography followed by dry etching. Manhattan-style Josephson junctions~\cite{muthusubramanian2024wafer} were patterned via electron beam lithography and subsequently constructed through double-angle evaporation with static oxidation to form the tunnel barrier~\cite{kreikebaum2020improving}. To suppress potential slotline modes and reduce signal crosstalk on the quantum chip, all control lines, including the qubits’ $XYZ$ lines and the DTCs’ $Z$ lines, are equipped with fully-capped tantalum airbridges~\cite{bu2025tantalum}. These airbridges were fabricated by depositing approximately 400 nm of tantalum metal onto scaffolds formed between two layers of photoresist, utilizing an aluminum sacrificial layer that was partially removed prior to deposition. Following deposition, the sample underwent an $80^{\circ}$C Remover PG bath for several hours, and the airbridges were finalized using a lift-off process.

\section{Experimental setup and extended data}

%--------------------------      Figure S2      --------------------------
\begin{figure*}[htb]
\includegraphics{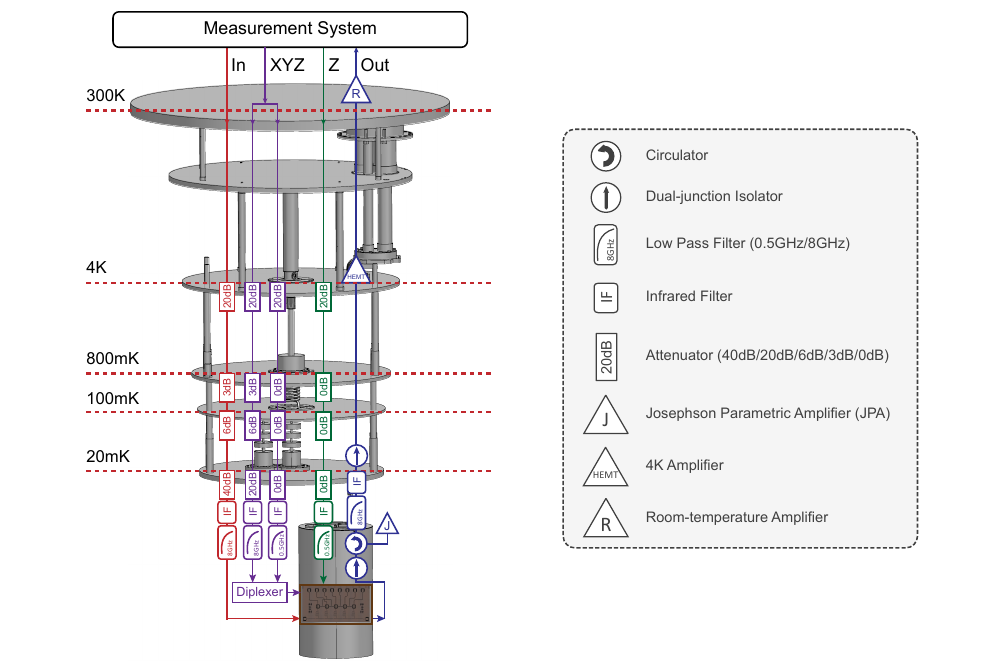}
\caption{\textbf{Measurement setup.} Red dashed lines indicate the temperatures at the dilution refrigerator’s various thermal stages. The readout input and output lines, the qubit $XYZ$ control lines, and the DTC $Z$ lines are distinguished by a color‑coding scheme (red, blue, purple, and green) for clarity.}
\label{fig:FigS2}
\end{figure*}
%--------------------------      Figure S2      --------------------------

In this section, we summarize the extended experimental data supporting this work.

\subsection{Device parameters}

\begin{table}[htb]
\caption{Device parameters. $\omega_r$ is the cavity frequencies for readout. $\omega_{j, \mathrm{max}}$, $\omega_{j, \mathrm{idle}}$, $\eta_j$ are the maximum frequencies, idle frequencies and anharmonicities for each qubit. $T_1$, $T_{2, \mathrm{echo}}$ are the corresponding energy relaxation time and echoed Ramsey dephasing time. Notice that the data was obtained during the experiment.}
%\begin{tabular}{cp{2.0cm}<{\centering}p{2.0cm}<{\centering}p{2.0cm}<{\centering}}
\begin{tabular*}{\linewidth}{@{\extracolsep{\fill}}cccc}
\hline
\hline
 &{$Q_1$} &{$Q_2$} &{$Q_3$} \tabularnewline
\hline
$\omega_r$ (GHz) &{$4.829$} &{$4.731$} &{$4.780$} \tabularnewline
$\omega_{j, \mathrm{max}}$ (GHz) &{$6.498$} &{$6.462$} &{$6.536$} \tabularnewline
$\omega_{j, \mathrm{idle}}$ (GHz) &{$6.343$} &{$6.433$} &{$6.033$} \tabularnewline
$\eta_j$ (MHz) &{$-221$} &{$-192$} &{$-218$} \tabularnewline
$T_1$ ($\mu$s) &{$55.6$} &{$25.4$} &{$47.1$} \tabularnewline
$T_{2, \mathrm{echo}}$ ($\mu$s) &{$9.4$} &{$23.4$} &{$1.3$} \tabularnewline
\hline
\end{tabular*}
\label{table:DeviceParameters}
\end{table}

\subsection{Measurement system}

The schematic of the measurement system is presented in Fig.~\ref{fig:FigS2}. Our five-qubit quantum chip is carefully packaged within an aluminum sample box and mounted inside infrared and magnetic shielding designed to preserve the qubits’ coherence times. Microwave signals are generated by custom-designed control electronics and routed through IQ mixers. These signals then pass through a series of attenuators and low-pass filters at successive temperature stages within the dilution refrigerator before reaching the qubits, enabling precise control and readout. Additionally, the $Z$-control signals for both the qubits and the DTCs are generated directly by the electronics without IQ mixing. Special attention is given to the selection of attenuators on the $Z$-lines to minimize thermal load on the dilution refrigerator, thereby preventing any adverse impact on its cooling performance. Since the qubit’s $XY$ and $Z$ lines are integrated into a common $XYZ$ line on the chip~\cite{manenti2021full}, the $XY$ and $Z$ signals are combined at the refrigerator’s base temperature stage using a diplexer. This three-port device merges low-frequency and high-frequency signals into a single output line while providing effective filtering at both frequency ports. The qubit readout output is initially amplified by a quantum-limited amplifier (IMPA) at the base temperature stage~\cite{roy2015broadband}, followed by further amplification via a high-electron-mobility transistor (HEMT) amplifier at the 4 K stage. Finally, the signal undergoes room-temperature amplification before being processed by the control electronics for data acquisition and analysis.

\subsection{DTC spectrum}

%--------------------------      Figure S3      --------------------------
\begin{figure}[tb]
\includegraphics{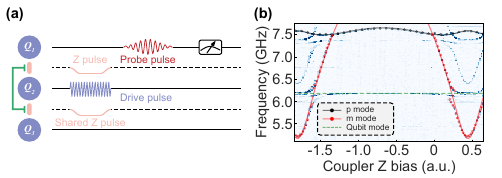}
\caption{\textbf{The DTC spectrum.} (a) Pulse sequence for measuring the spectrum of DTC$_1$, which couples qubits $Q_1$ and $Q_2$. A similar sequence is used for DTC$_2$, which couples $Q_2$ and $Q_3$. (b) Measured spectrum of DTC$_1$. The $p$-mode and $m$-mode are fitted with a black dotted line and a red dotted line, while the qubit mode is indicated by a green dashed line.}
\label{fig:FigS3}
\end{figure}
%--------------------------      Figure S3      --------------------------

Characterization of the DTC spectrum is essential for accurately determining its fundamental properties, including the frequency ranges of the $p$-mode and $m$-mode, the coupling between these modes, and \blue{the coupling} between the $m$-mode and the qubits. Since the DTCs lack dedicated readout resonators, direct dispersive readout to scan their full spectrum is not feasible. Therefore, an indirect measurement approach is employed, leveraging the coupling between the DTCs and the qubits to achieve precise spectral measurement.

In this experiment, the $Z$ control lines of DTC$_1$ (coupling qubits $Q_1$ and $Q_2$) and DTC$_2$ (coupling qubits $Q_2$ and $Q_3$) are shared. Consequently, it is necessary to select appropriate qubits for spectral measurement of each DTC. Specifically, $Q_2$ cannot be used for characterizing the spectrum of either DTC, as it is coupled to both, resulting in a combined spectral response. For instance, when scanning the spectrum of DTC$_1$, $Q_1$ should be used as the measurement qubit; similarly, $Q_3$ should be used for DTC$_2$. The experimental pulse sequence for measuring the spectrum of DTC$_1$ is illustrated in Fig.~\ref{fig:FigS3}(a). A $Z$ control pulse is applied to modulate the frequency of DTC$_1$ (noting that DTC$_2$ is also affected due to the shared $Z$ line). A drive pulse is simultaneously applied via the control line of $Q_2$ to excite DTC$_1$ through microwave crosstalk interaction (since the DTC lacks an $XY$ drive line), followed by a probe pulse on $Q_1$ to acquire the full spectrum, as shown in Fig.~\ref{fig:FigS3}(b).

As the external magnetic flux applied to the DTC varies, both the $m$-mode and $p$-mode frequencies shift from their maximum toward minimum values. Fitting the data reveals that the $m$-mode maximum frequency is approximately 9.0 GHz with a minimum near 5.1 GHz, while the $p$-mode maximum frequency is around 7.6 GHz with a minimum near 7.4 GHz. The $m$-mode exhibits a significantly faster frequency tuning rate. Around 7.5 GHz, the $m$-mode crosses the $p$-mode, resulting in an avoided crossing. As the $m$-mode frequency further decreases to approximately 6.1 GHz, it resonates with the qubit mode. Due to the IQ-mixing-based electronic control system employed, sideband signals appear in the measurement; these are also indicated in Fig.~\ref{fig:FigS3}(b), demonstrating that both the $m$-mode and qubit mode signals exhibit corresponding sideband features. From the measured DTC spectrum, we identify the approximate coupling “off” point. By biasing the DTC near the point and performing \blue{a} fine scan of the $ZZ$ coupling between qubits, we obtain the data presented in Fig.~2(e) of the main text, which enables determination of the optimal idle position for the DTC. During the implementation of CZ and parametric gates, the DTC spectrum further informs the selection of the flux bias to achieve the desired coupling strength, as demonstrated in Fig.~5(a) of the main text.

\subsection{$Z$ pulse distortion}

In a frequency-tunable transmon system incorporating STCs or DTCs, $Z$ pulses are used to implement both single- and two-qubit gates. Precise control of these pulses is therefore critical—for example, for CZ and parametric two-qubit gates as well as single-qubit operations such as fast reset. In practice, however, $Z$ pulses are often distorted by limitations in the room-temperature electronics, microwave components in the cabling, the sample-box PCB, and wire bonds. These distortions are most pronounced on the pulse rising and falling edges and lead to waveform deviations at the device that cause unwanted frequency shifts and leakage.

\begin{figure}[tb]
\includegraphics{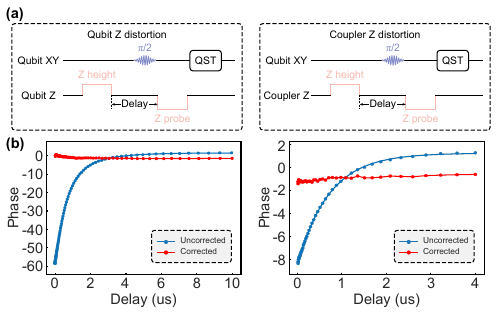}
\caption{\textbf{$Z$ pulse distortion.} (a) Pulse sequences for measuring the $Z$ pulse distortion of qubit (left panel) and DTC (right panel). (b) The extracted phase variation with uncorrected and corrected $Z$ pulse distortion for qubit (left panel) and DTC (right panel).}
\label{fig:FigS4}
\end{figure}

To calibrate and mitigate $Z$ pulse distortion for both the qubit and the DTC, we treat each device as a low-temperature oscilloscope to directly characterize the waveform it experiences~\cite{foxen2018high, rol2020time}, and then apply data-driven compensation. The pulse sequences for measuring $Z$ distortion are shown in Fig.~\ref{fig:FigS4}(a). For the qubit, we probe a flux region where the qubit frequency is particularly sensitive to $Z$ amplitude so that small distortions produce a measurable phase shift. An initial $Z$ waveform is applied to induce the distortion; by varying the delay between this waveform and a subsequent probe $Z$ pulse and using a Ramsey sequence, we measure the qubit phase evolution and fit the data to quantify the distortion. We then correct the qubit $Z$ pulse according to the fitted parameters and validate the correction by repeating the measurement and confirming suppression of the phase dependence on delay. The DTC calibration follows the same principle but requires an indirect readout because the DTC lacks an independent readout resonator. Owing to the strong qubit–DTC coupling, we infer the DTC response via the qubit by probing a region near the qubit–DTC coupling where the DTC’s sensitivity to $Z$ pulses is enhanced. Using the same Ramsey-based fitting procedure, we extract the DTC distortion and its tail (decay) coefficients. 

After obtaining correction parameters for both devices, we apply waveform compensation to correct short- and long-time distortion components. The compensated $Z$ pulses show substantially improved fidelity for $Z$-pulse-based single- and two-qubit operations. The uncorrected and corrected pulse performance is summarized in Fig.~\ref{fig:FigS4}(b).

%--------------------------     Figure S4      --------------------------
%--------------------------      Figure S4      --------------------------

\section{Numerical simulations}

%--------------------------      Figure S5      --------------------------
\begin{figure}[tb]
\includegraphics{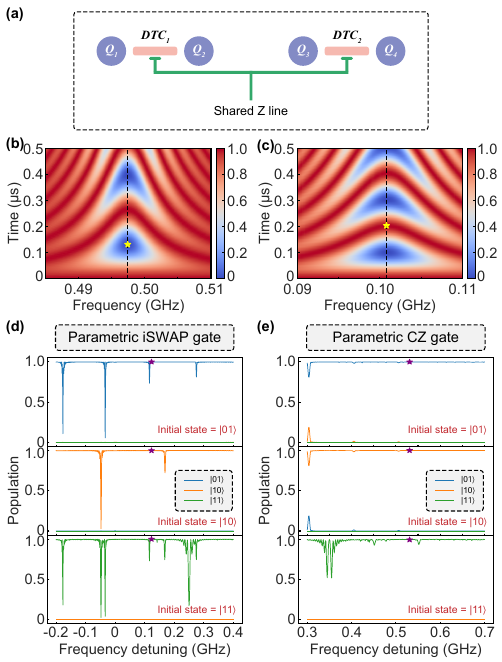}
\caption{\textbf{Numerical simulations on quantum state evolution of spectator qubit pair.} (a) Schematic diagram of the multiplexing structure for coupler control lines: $Q_1$ and $Q_2$ are coupled via DTC$_1$, while $Q_3$ and $Q_4$ are coupled via DTC$_2$. The two tunable couplers (DTC$_1$ and DTC$_2$) are connected via a shared $Z$-line to realize control line multiplexing.  (b), (c) Dynamics of representative initial states used to implement the parametric (b) iSWAP and (c) CZ gates. The chevron patterns are the evolution of $\ket{01}$ for the parametric iSWAP gate and $\ket{11}$ for the parametric CZ gate, with the pulse-parameter sets indicated by yellow stars yielding process fidelities of 99.993\% (iSWAP) and 99.992\% (CZ), respectively. (d), (e) Evolution of spectator qubit pair initialized in $\ket{01}$, $\ket{10}$, and $\ket{11}$ as a function of qubit detuning by executing the parametric (d) iSWAP and (e) CZ gates. The pulses are consistent with those in (b) and (c), respectively. The process fidelities corresponding to the parameter sets indicated by purple stars are 99.987\% and 99.955\%.}
\label{fig:FigS5}
\end{figure}
%--------------------------      Figure S5      --------------------------

Parametric gates can facilitate scalable deployment of DTC-based couplers. In this section, we further present numerical simulations of parametric gates based on the multiplexed DTC scheme. 

The simulated model layout is shown in Fig.~\ref{fig:FigS5}(a): two DTCs couple a left qubit pair ($Q_1$ and $Q_2$) and a right qubit pair ($Q_3$ and $Q_4$), and both DTCs are driven by a common $Z$ control line. The left and right pairs are assumed sufficiently separated to neglect direct spurious coupling (scalable applications can be found in the right panel in Fig.~5(d) in the main text). The pair on which parametric gates are applied is called the target qubit pair; the other is the spectator qubit pair. The Hamiltonian for each pair is given in Eq.~(1) in the main text. 

We first simulate a parametric iSWAP gate implemented via the multiplexed DTC with the target qubit pair initialized in state $\ket{01}$. The applied parametric pulse is designed according to the DTC’s effective $XY$ coupling so that the effective qubit–qubit coupling oscillates sinusoidally around zero. By setting the drive frequency close to the qubit detuning, we obtain coherent population exchange as illustrated in Fig.~\ref{fig:FigS5}(b). The optimal operating points are marked by yellow stars and quantum process tomography (QPT) is applied at this point to measure the gate fidelity, yielding a fidelity of over 99.992\%. 

However, in the multiplexed configuration, periodic modulation of the shared $Z$ control line of both DTCs also perturbs the spectator qubit pair. To quantify this unwanted crosstalk, we set the initial states of the spectator qubit pair to $\ket{01}$, $\ket{10}$, and $\ket{11}$, respectively, and then modulation pulses required for gate control are applied to the $Z$ line to implement the iSWAP gate operation on the target qubit pair. After the implementation of the iSWAP gate, we focus on observing the initial states of the spectator qubit pair. The results are shown in Fig.~\ref{fig:FigS5}(d) (upper panel: initial state $\ket{01}$; middle panel: initial state $\ket{10}$; bottom panel: initial state $\ket{11}$). It can be found that, at certain detunings, significant deviations occur due to leakage into the coupler’s $m$-mode when that mode is driven; at detunings corresponding to the star markers, this gate procedure approximates the identity operation ($I\otimes I$) and QPT yields a fidelity of 99.987\%. These observations demonstrate strong crosstalk suppression in the multiplexed coupler configuration. Further improvements are expected through coupler-parameter optimization: reducing gate duration and adjusting coupler parameters can suppress spurious excitation of non‑target modes (e.g., the $m$‑mode) and thereby lower the probability of state leakage.

The parametric CZ gate simulations follow a similar procedure. We initialize the target qubit pair in state $\ket{11}$, and apply a parametric tone tuned to the appropriate resonance that couples the $\ket{11}$ computational manifold to the relevant noncomputational state. Under these conditions the $\ket{11}$ population exhibits periodic oscillations, as shown in Fig.~\ref{fig:FigS5}(c); the yellow star markers denote the optimal operating point, reaching up to 99.993\% fidelity. Meanwhile, crosstalk for the parametric CZ gate is weaker: across a wide detuning range the spectator qubit pair state populations remain above 99.9\%, and QPT at the marked star point yields a fidelity of 99.955\%, as depicted in Fig.~\ref{fig:FigS5}(e) (upper panel: initial state $\ket{01}$; middle panel: initial state $\ket{10}$; bottom panel: initial state $\ket{11}$). 

These numerical results demonstrate that high‑fidelity two-qubit parametric gates (iSWAP and CZ) can be realized based on our multiplexed DTC scheme. The simulations establish the feasibility of this approach and provide a basis for further experimental optimization and scaling studies.

%%\bibliographystyle{naturemag_nourl}
%\makeatletter
%\renewcommand\@biblabel[1]{#1.}
%\makeatother
%%\bibliography{Citation_sp}
%
%\bibliographystyle{unsrt}
%\bibliography{Citation}

\clearpage

\bigskip